\documentclass[a4paper]{jpconf}
\usepackage{graphicx,color,latexsym}
\begin{document}
\title{Hydrodynamic representation of the Klein-Gordon-Einstein equations in
the weak field limit}

\author{Abril Su\'arez and Pierre-Henri Chavanis}

\address{Laboratoire de Physique Th\'eorique, Universit\'e Paul
Sabatier, 118 route de Narbonne 31062 Toulouse, France}

\ead{suarez@irsamc.ups-tlse.fr, chavanis@irsamc.ups-tlse.fr}

\begin{abstract}
Using a generalization of the Madelung transformation, we derive the
hydrodynamic representation of the Klein-Gordon-Einstein equations in the weak
field limit. We consider a complex self-interacting scalar field with an
arbitrary potential of the form $V(|\varphi|^2)$.  We compare the results with
simplified models in which the gravitational potential is introduced by hand in
the Klein-Gordon equation, and assumed to satisfy a (generalized) Poisson
equation.  Nonrelativistic hydrodynamic equations based on the
Schr\"odinger-Poisson equations or on the Gross-Pitaevskii-Poisson equations are
recovered in the limit $c\rightarrow +\infty$.
\end{abstract}

\section{Introduction}
\label{sec_intro}

Scalar fields (SF) play an important role in particle physics,
astrophysics, and cosmology \cite{zee}. Their evolution
is usually described by the Klein-Gordon (KG) equation
\cite{klein1,gordon} which can be
viewed as a relativistic extension of the Schr\"odinger equation
\cite{schrodinger1}. The KG equation describes spin-$0$ 
particles (bosons) that can be charged (complex SF) or neutral (real
SF).
The coupling between the KG equation and
gravity through
the
Einstein equations,  leading to the
Klein-Gordon-Einstein (KGE) equations, 
was first considered in the context of boson
stars
\cite{kaup,rb,colpi}. It has also been proposed that dark matter (DM) halos may
be made of a SF
described by the KGE equations (see,
e.g., \cite{revueabril,revueshapiro,bookspringer} for recent reviews). Actually,
at the galactic scale, the Newtonian limit is valid so that DM halos can be
described by the Schr\"odinger-Poisson (SP) equations or by the
Gross-Pitaevskii-Poisson (GPP) equations. In that case, the wave function
$\psi(\vec
x,t)$ describes a Bose-Einstein condensate (BEC) at $T=0$, and the
self-interaction of the bosons is measured by their scattering length $a_s$.
Therefore, DM halos could be gigantic quantum objects made of BECs. The wave
properties of bosonic DM may
stabilize the system against gravitational collapse, providing halo cores
and sharply suppressing small-scale linear power. This may
solve the
problems of the cold dark matter (CDM) model such as the cusp problem and the
missing
satellite
problem. The  scalar
field dark matter (SFDM) model and the BEC dark matter (BECDM) model, also
called $\Psi$DM models,
have received much attention in the last years.

Since DM may be a SF, it is of
considerable interest to study the cosmological implications of this scenario.
The cosmological evolution of a spatially
homogeneous noninteracting real SF described by the KGE equations competing
with baryonic matter, radiation and dark energy was considered by
Matos {\it et al.} \cite{mvm}. They found that real SFs display fast
oscillations but that, on the mean, they reproduce the cosmological
predictions of the standard
$\Lambda$CDM model. The study of
perturbations was considered by Su\'arez and
Matos
\cite{abrilMNRAS} for a self-interacting real SF
described by the Klein-Gordon-Poisson (KGP) equations and by
Maga\~na {\it et al.} \cite{abrilJCAP} for a noninteracting
real SF described by the KGE equations. These studies show that the
perturbations can grow in the linear regime, leading, in the nonlinear regime,
to the formation of structures corresponding to DM halos. This is in
agreement with the early work of Khlopov {\it et al.} \cite{khlopov} who studied
the Jeans instability of a relativistic SF in a static
background. The case of a complex self-interacting SF representing BECDM was
considered by
Chavanis \cite{chavaniscosmo} in the context of
Newtonian cosmology. His study is based on the GPP equations. Harko
\cite{harkocosmo} and Chavanis \cite{chavaniscosmo} developed an approximate
relativistic BEC cosmology and found that the perturbations grow faster in BECDM
as compared to
$\Lambda$CDM. Recently, Li {\it et al.} \cite{shapiro} developed an exact
relativistic
cosmology
for a complex self-interacting SF/BEC based on the KGE equations. They studied 
the
evolution of the homogeneous background and showed that the Universe
undergoes three successive phases: a stiff matter era, followed by a radiation
era due to the SF (that exists only for self-interacting SFs), and finally a
matter era similar to CDM.

Instead of working directly in terms of a SF, we can adopt a
fluid approach and work with hydrodynamic equations. In the case of the
Schr\"odinger equation, this hydrodynamic approach was introduced by
Madelung  \cite{madelung}. 
He showed that the Schr\"odinger equation is equivalent to the Euler equations
for an irrotational fluid with an additional quantum
potential
arising from the finite value of $\hbar$.   This
hydrodynamic representation has been used by B\"ohmer and Harko \cite{bohmer}
and by Chavanis \cite{chavaniscosmo,prd1,prd2} among others in the case of BECDM
and in
the case of BEC stars
\cite{chavharko}. This hydrodynamic approach has been generalized by Su\'arez
and Matos
\cite{abrilMNRAS,smz}
in the context of the KG equation. They used it to
study the formation of structures in the Universe, assuming that DM is
in the form of a fundamental SF with a $\lambda \varphi^4$ potential.

In the works \cite{abrilMNRAS,smz}, the SF is taken to be real and the
gravitational
potential is
introduced by hand in the KG equation, and assumed to be determined by the
classical 
Poisson equation where the source is the rest-mass density $\rho$. This leads to
the
KGP equations. However, this treatment is not self-consistent since it combines
relativistic and nonrelativistic equations. In these
Proceedings, we derive the hydrodynamic
representation of a complex SF coupled to gravity through
the Einstein
equations in the weak field approximation. In this way, we develop a
self-consistent relativistic treatment. Throughout this work, we use the
conformal Newtonian gauge which takes into account metric perturbations up to
first order.
We  consider only scalar perturbations. This is sufficient if we are interested
in calculating
observational consequences of the SF dynamics in the linear regime.
We compare our results with those
obtained from the heuristic KGP equations. In these Proceedings, we
develop the basic formalism and provide the main equations for an arbitrary
self-interaction potential of the form $V(|\varphi|^2)$. Details and
applications of our formalism can be found in our research
papers \cite{abrilph1,abrilph2}.

\section{The  conformal Newtonian gauge}

We consider the  KGE equations in the weak field limit $\Phi/c^2\ll
1$. The equations that we derive are valid at the order $O(\Phi/c^2)$.
We work with the conformal Newtonian gauge which is a perturbed form
of the Friedmann-Lema\^itre-Robertson-Walker  (FLRW) line element \cite{ma}. We
consider the simplest form of the
Newtonian gauge, only taking
into account scalar perturbations which are the ones that contribute to
the formation of structures in cosmology. We neglect anisotropic
stresses. We assume that the Universe is flat in
agreement with the observations of the cosmic microwave background (CMB).
Under these conditions, our line element is given by
\begin{equation}
ds^2=c^2\left(1+2\frac{\Phi}{c^2}\right)dt^2-a(t)^2\left(1-2\frac{\Phi}{c^2}
\right)\delta_{ij}dx^idx^j,
\label{kge2}
\end{equation}
where ${\Phi}/{c^2}\ll 1$. In this metric, $\Phi(\vec x,t)$ represents the
gravitational potential of
classical Newtonian gravity and $a(t)$ is the (dimensionless)
scale factor. 

\section{The Lagrangian of the scalar field}

We assume that DM can be described by a  complex
SF which is a continuous
function of space and time defined at each point by
$\varphi(x^\mu)=\varphi(x,y,z,t)$. The action of the relativistic SF
is
\begin{equation}
S_\varphi=\int d^4x\sqrt{-g}\mathcal{L}_\varphi,
\label{tb1}
\end{equation}
where $\mathcal{L}_\varphi=\mathcal{L}_\varphi(\varphi,
\varphi^*,\partial_\mu\varphi,\partial_\mu\varphi^*)$
is the Lagrangian density and $g={\rm det}(g_{\mu\nu})$ is the determinant of
the metric tensor. We adopt the following generic Lagrangian density 
\begin{eqnarray}
{\cal L}_\varphi=\frac{1}{2}g^{\mu\nu}\partial_{\mu}\varphi^*
\partial_{\nu}\varphi-\frac{m^2c^2}{2\hbar^2}|\varphi|^2-V(|\varphi|^2)
\label{tb2}
\end{eqnarray}
which is written for a metric signature $(+,-,-,-)$.  The second (quadratic)
term is the rest-mass term. The potential $V(|\varphi|^2)$ takes into account
the self-interaction of the SF. In certain applications, it is
relevant to
consider a quartic potential of the form
\begin{equation}
V(|\varphi|^2)=\frac{m^2}{2\hbar^4}\lambda|\varphi|^4,
\label{tb3}
\end{equation} 
where $\lambda$ is the
self-interaction coupling strength. If the SF describes a
BEC at $T=0$, the quartic potential (\ref{tb3})  models the two-particle
self-interaction.\footnote{It is a good approximation to ignore
higher-order
interactions when the boson gas is dilute, i.e., when the particle
self-interaction range is much smaller than the mean interparticle distance.}
In
that case, $m$ represents the mass of the bosons and $\lambda$ is a constant
related to the s-wave scattering length of the bosons $a_s$ (measured in the
center of mass frame) by $\lambda=4\pi
a_s\hbar^2/m$. This corresponds to the first Born approximation. The potential
can then be written as 
\begin{equation}
V(|\varphi|^2)=\frac{2\pi a_s
m}{\hbar^2}|\varphi|^4.
\label{tb4}
\end{equation}

\section{The energy-momentum tensor}

The energy-momentum tensor of the SF is
\begin{eqnarray}
T_{\mu\nu}=2\frac{\partial\mathcal{L}_\varphi}{\partial
g^{\mu\nu}}-g_{\mu\nu}\mathcal{L}_{\varphi}.
\label{tb11}
\end{eqnarray}
For the generic Lagrangian (\ref{tb2}), it takes the form
\begin{eqnarray}
T_{\mu\nu}=\frac{1}{2}(\partial_{\mu}\varphi^*
\partial_{\nu}\varphi+\partial_{\nu}\varphi^* \partial_{\mu}\varphi)
-g_{\mu\nu}\left
\lbrack\frac{1}{2}g^{\rho\sigma}\partial_{\rho}\varphi^*\partial_{\sigma}
\varphi-\frac{m^2c^2}{2\hbar^2}|\varphi|^2-V(|\varphi|^2)\right \rbrack.
\label{tb12}
\end{eqnarray}
By analogy with the energy-momentum tensor of a perfect fluid, the energy
density and the pressure tensor of the SF are defined by $\epsilon=T_0^0$ and
$P_i^j=-T_i^j$. The conservation of
the energy-momentum tensor, which
results from the  Noether theorem, writes  $D_{\nu}T^{\mu\nu}=0$.

\section{The Klein-Gordon equation}

The equation of motion for the SF can be obtained from the principle
of least action. Imposing  $\delta S_{\varphi}=0$ for arbitrary variations 
$\delta\varphi$
and $\delta\varphi^*$, we obtain the Euler-Lagrange
equation  
\begin{eqnarray}
D_{\mu}\left\lbrack\frac{\partial {\cal L}_\varphi}{\partial
(\partial_{\mu}\varphi)^*}\right\rbrack-\frac{\partial {\cal
L}_\varphi}{\partial\varphi^*}=0,
\label{tb5}
\end{eqnarray}
where $D$ is the covariant derivative. For the generic Lagrangian (\ref{tb2}),
 this leads to the KG
equation
\begin{equation}
\Box\varphi+\frac{m^2c^2}{\hbar^2}\varphi+2V(|\varphi|^2),_{\varphi^*}=0,
\label{tb6}
\end{equation}
where $\Box$ is the d'Alembertian operator
\begin{equation}
\Box\equiv D_{\mu}(g^{\mu\nu}\partial_{\nu})=\frac{1}{\sqrt{-g}}
\partial_\mu(\sqrt{-g}\, g^{\mu\nu}\partial_\nu)
\label{tb7}
\end{equation}
and
\begin{equation}
V(|\varphi|^2),_{\varphi^*}=\frac{dV}{d|\varphi|^2}\varphi.
\label{tb6w}
\end{equation}
Computing the d'Alembertian (\ref{tb7}) with the Newtonian gauge, we obtain the
KG equation
\begin{equation}
\frac{1}{c^2}\frac{\partial^2\varphi}{\partial
t^2}+\frac{3H}{c^2}\frac{\partial\varphi}{\partial
t}-\frac{1}{a^2}\left(1+\frac{4\Phi}{c^2}\right)\Delta\varphi
-\frac{4}{c^4}\frac{\partial\Phi}{
\partial t}\frac{\partial\varphi}{\partial
t}+\left (1+\frac{2\Phi}{c^2}\right) \frac{m^2
c^2}{\hbar^2}\varphi
+2\left(1+2\frac{\Phi}{c^2}\right)V,_{\varphi^*} =0,\label{kge4}
\end{equation}
where $H=\dot a/a$ is the Hubble constant. Using the expression (\ref{tb12})  of
the energy-momentum tensor, the energy
density and
the pressure are given by 
\begin{eqnarray}
\epsilon=T_0^0=\frac{1}{2c^2}\left (1-\frac{2\Phi}{c^2}\right
)\left
|\frac{\partial\varphi}{\partial t}\right |^2 
+\frac{1}{2a^2}\left (1+\frac{2\Phi}{c^2}\right
)|\vec\nabla\varphi|^2+\frac{m^2c^2}{2\hbar^2}|\varphi|^2+V(|\varphi|^2),
\label{kge5}
\end{eqnarray}
\begin{eqnarray}
P=-\frac{1}{3}(T_1^1+T_2^2+T_3^3)=\frac{1}{2c^2}\left (1-\frac{2\Phi}{c^2}\right
)\left
|\frac{\partial\varphi}{\partial t}\right |^2 
-\frac{1}{6a^2}\left (1+\frac{2\Phi}{c^2}\right
)|\vec\nabla\varphi|^2-\frac{m^2c^2}{2\hbar^2}
|\varphi|^2-V(|\varphi|^2).\nonumber\\
\label{kge6}
\end{eqnarray}

\section{The Einstein equations}

The Einstein equations write
\begin{equation}
R_{\mu\nu}-\frac{1}{2}g_{\mu\nu}R=\frac{8\pi G}{c^4}T_{\mu\nu},
\label{kge9}
\end{equation}
where $R_{\mu\nu}$ is the Ricci tensor,  $R$ is the Ricci scalar, and $G$ is
Newton's gravitational constant \cite{weinberg}. The conservation of
the energy-momentum tensor is automatically included in
the Einstein equations. The time-time component of the Einstein equations is 
\begin{eqnarray}
R_0^0-\frac{1}{2}R=\frac{8\pi G}{c^4}T_0^0.
\label{kge10}
\end{eqnarray}
With the Newtonian gauge, it can be written as
\begin{eqnarray}
\frac{\Delta\Phi}{4\pi Ga^2}=\frac{\epsilon}{c^2}-\frac{3H^2}{8\pi
G}+\frac{3H}{4\pi G
c^2}\left (\frac{\partial\Phi}{\partial t}+H\Phi\right ).
\label{kge11b}
\end{eqnarray}
Using the expression (\ref{kge5}) of the energy density $\epsilon$,
which represents the time-time component of the energy-momentum tensor, we get
\begin{eqnarray}
\frac{\Delta\Phi}{4\pi G a^2}=\frac{1}{2c^4}\left (1-\frac{2\Phi}{c^2}\right
)\left |\frac{\partial\varphi}{\partial t}\right |^2
+\frac{1}{2a^2c^2}\left (1+\frac{2\Phi}{c^2}\right
)|\vec\nabla\varphi|^2+\frac{m^2}{2\hbar^2}|\varphi|^2\nonumber\\
+\frac{1}{c^2}V(|\varphi|^2)-\frac{3H^2}{8\pi G}+\frac{3H}{4\pi G
c^2}\left (\frac{\partial\Phi}{\partial t}+H\Phi\right ).
\label{kge12}
\end{eqnarray}
Eqs. (\ref{kge4}) and (\ref{kge12}) form the KGE equations.

\section{Spatially homogeneous scalar field}

For a spatially homogeneous SF with $\varphi_b(\vec
x,t)=\varphi_b(t)$ and $\Phi_b(\vec
x,t)=0$, the KG equation (\ref{kge4}) reduces to
\begin{eqnarray}
\frac{1}{c^2}\frac{d^2\varphi_b}{dt^2}+\frac{3H}{c^2}\frac{d\varphi_b}{dt}+\frac
{m^2
c^2}{\hbar^2}\varphi_b
+2V,_{\varphi_b^*
} =0.
\label{kgp6b}
\end{eqnarray}
In that case, the energy-momentum tensor is diagonal and isotropic,
$T^{\mu}_{\nu}={\rm diag} (\epsilon_b,-P_b,-P_b,-P_b)$. The energy density
$\epsilon_b(t)$ and the pressure $P_b(t)$ are given by
\begin{equation}
\epsilon_b=\frac{1}{2c^2}\left |\frac{d\varphi_b}{d
t}\right|^2+\frac{m^2c^2}{2\hbar^2}|\varphi|^2+V(|\varphi_b|^2),\qquad
P_b=\frac{1}{2c^2}\left |\frac{d\varphi_b}{d
t}\right|^2-\frac{m^2c^2}{2\hbar^2}|\varphi|^2-V(|\varphi_b|^2).
\label{kgp7b}
\end{equation}
From these equations, we obtain the continuity equation
\begin{equation}
\frac{d\epsilon_b}{dt}+3H(\epsilon_b+P_b)=0
\label{frid1}
\end{equation}
which is one of the Friedmann equations \cite{weinberg}. The other Friedmann
equation is
obtained from the Einstein equation (\ref{kge12}) that reduces to
\begin{eqnarray}
H^2=\frac{8\pi G}{3c^2}\epsilon_b.
\label{kge12b}
\end{eqnarray}
This relation shows that the term $-3H^2/8\pi G=-\epsilon_b$ in the Einstein
equation (\ref{kge11b}) plays the role of a neutralizing
background.\footnote{In a static universe, the source of the
gravitational potential is $\epsilon/c^2$. In an expanding universe,  the
source of
the gravitational
potential is $(\epsilon-\epsilon_b)/c^2$, so that $\Phi=0$ when
$\epsilon=\epsilon_b$. When we work in the comoving frame, the
expansion of the
universe amounts to
subtracting a neutralizing background $-\epsilon_b/c^2$ to the density
$\epsilon/c^2$ \cite{peebles}, as in the Jellium model of
plasma physics. This is the correct way to solve the Jeans
swindle \cite{bt}.}
From Eqs. (\ref{frid1}) and (\ref{kge12b}), we easily obtain
\begin{eqnarray}
\frac{\ddot a}{a}=-\frac{4\pi G}{3c^2}(\epsilon_b+3P_b).
\label{kge12bw}
\end{eqnarray}

\section{The Gross-Pitaevskii-Einstein equations}

The KG equation without self-interaction can be viewed as a
relativistic generalization of the Schr\"odinger equation. Similarly, the
KG equation with a self-interaction can be viewed as a
relativistic generalization of the GP equation. In
order to recover the  Schr\"odinger and GP equations in the
nonrelativistic limit $c\rightarrow +\infty$, we make the transformation
\cite{zee}:
\begin{eqnarray}
\varphi(\vec{x},t)=\frac{\hbar}{m}e^{-i m c^2 t/\hbar}\psi(\vec{x},t).
\label{kgp12}
\end{eqnarray}
The prefactor $\hbar/m$ is
justified in \ref{sec_cons}. Mathematically, we can
always make this change of variables. 
However, we emphasize that it is only in the nonrelativistic limit $c\rightarrow
+\infty$ that $\psi$ has the interpretation of a wave function, and that
$|\psi|^2=\rho$ has the interpretation of a rest-mass density. In the
relativistic regime, $\psi$ and $\rho=|\psi|^2$ do not have a clear
physical
interpretation. We will call them ``pseudo wave function'' and ``pseudo
rest-mass
density''. Nevertheless, it is perfectly legitimate to work with these
variables and, as we shall see,  the equations written in terms of these
quantities take relatively simple forms that generalize naturally the
nonrelativistic
ones.

Substituting Eq.
(\ref{kgp12}) into Eqs. (\ref{kge4}) and
(\ref{kge12}), we obtain 
\begin{eqnarray}
i\hbar\frac{\partial\psi}{\partial t}-\frac{\hbar^2}{2m
c^2}\frac{\partial^2\psi}{\partial t^2}-\frac{3}{2}H\frac{\hbar^2}{m
c^2}\frac{\partial\psi}{\partial t}
+\frac{\hbar^2}{2 m a^2}\left
(1+\frac{4\Phi}{c^2}\right )\Delta\psi-m\Phi \psi\nonumber\\
-\left
(1+\frac{2\Phi}{c^2}\right )m \frac{dV}{d|\psi|^2}\psi+\frac{3}{2}i\hbar
H\psi
+\frac{2\hbar^2}{m c^4}\frac{\partial\Phi}{\partial t}\left
(\frac{\partial \psi}{\partial t}-\frac{i m c^2}{\hbar}\psi\right )=0,
\label{kge13}
\end{eqnarray}
\begin{eqnarray}
\frac{\Delta\Phi}{4\pi G a^2}=\left (1-\frac{\Phi}{c^2}\right
)|\psi|^2
+\frac{\hbar^2}{2m^2c^4}\left (1-\frac{2\Phi}{c^2}\right )\left
|\frac{\partial\psi}{\partial t}\right |^2+\frac{\hbar^2}{2a^2m^2c^2}\left
(1+\frac{2\Phi}{c^2}\right )|\vec\nabla\psi|^2\nonumber\\
+\frac{1}{c^2}V(|\psi|^2)-\frac{\hbar}{m c^2}\left
(1-\frac{2\Phi}{c^2}\right ){\rm Im} \left (\frac{\partial\psi}{\partial
t}\psi^*\right ) -\frac{3H^2}{8\pi G}+\frac{3H}{4\pi G c^2}\left
(\frac{\partial\Phi}{\partial t}+H\Phi\right ).
\label{kge14}
\end{eqnarray}
Eq. (\ref{kge13})
can be interpreted as a generalized
Schr\"odinger equation (in the absence of self-interaction) or
as a generalized GP equation (in the presence of self-interaction).
It is coupled to the
Einstein equation (\ref{kge14}). The energy density and the pressure can be
written as
\begin{eqnarray}
\epsilon=\frac{\hbar^2}{2m^2c^2}\left (1-\frac{2\Phi}{c^2}\right )\left
|\frac{\partial\psi}{\partial t}\right |^2
-\frac{\hbar}{m}\left
(1-\frac{2\Phi}{c^2}\right ){\rm Im} \left
(\frac{\partial\psi}{\partial
t}\psi^*\right )\nonumber\\
+\frac{\hbar^2}{2a^2m^2}\left (1+\frac{2\Phi}{c^2}\right
)|\vec\nabla\psi|^2
+\frac{1}{2}c^2\left (1-\frac{2\Phi}{c^2}\right
)|\psi|^2+\frac{1}{2}c^2|\psi|^2+V(|\psi|^2),
\label{kgp14b}
\end{eqnarray}
\begin{eqnarray}
P=\frac{\hbar^2}{2m^2c^2}\left (1-\frac{2\Phi}{c^2}\right )\left
|\frac{\partial\psi}{\partial t}\right |^2
-\frac{\hbar}{m}\left
(1-\frac{2\Phi}{c^2}\right ){\rm Im} \left
(\frac{\partial\psi}{\partial
t}\psi^*\right )\nonumber\\
-\frac{\hbar^2}{6a^2m^2}\left (1+\frac{2\Phi}{c^2}\right
)|\vec\nabla\psi|^2
+\frac{1}{2}c^2\left (1-\frac{2\Phi}{c^2}\right
)|\psi|^2-\frac{1}{2}c^2|\psi|^2-V(|\psi|^2).
\label{kgp14c}
\end{eqnarray}
Eqs. (\ref{kge13}) and (\ref{kge14}) form the  GPE equations. In the
nonrelativistic limit $c\rightarrow +\infty$, they reduce to the GPP
equations \cite{chavaniscosmo}:
\begin{eqnarray}
i\hbar\frac{\partial\psi}{\partial
t}+\frac{3}{2}i\hbar
H\psi=-\frac{\hbar^2}{2 m a^2}\Delta\psi+m\Phi \psi
+m\frac{dV}{d|\psi|^2}\psi,
\label{nr1}
\end{eqnarray}
\begin{eqnarray}
\frac{\Delta\Phi}{4\pi G a^2}=|\psi|^2 -\frac{3H^2}{8\pi G}.
\label{nr2}
\end{eqnarray}
For the quartic potential (\ref{tb4}), we have
\begin{equation}
V(|\psi|^2)=\frac{2\pi a_s\hbar^2}{m^3}|\psi|^4.
\label{tb4w}
\end{equation}

\section{The hydrodynamic representation}

Important characteristics of the system are revealed  by reformulating the
KGE equations in the form of hydrodynamic equations. This can
be done
at the level of the GPE equations
(\ref{kge13})-(\ref{kge14}) via the Madelung  transformation \cite{madelung}. To
that purpose, we write the pseudo wave function $\psi$ as
\begin{eqnarray}
\psi(\vec{x},t)=\sqrt{\rho(\vec{x},t)} e^{iS(\vec{x},t)/\hbar},\label{kgp15}
\end{eqnarray}
where $\rho=|\psi|^2$ plays the role of a pseudo rest-mass density and $S$ plays
the role of a pseudo action. Following
Madelung, we also define a pseudo velocity field as
\begin{eqnarray}
\vec{v}(\vec{x},t)=\frac{\vec\nabla S}{ma},
\label{kgp16}
\end{eqnarray}
where the scale factor $a$ has been introduced in order to
take into account the expansion of the Universe. We
note that this velocity field is irrotational.

Substituting Eqs. (\ref{kgp15})-(\ref{kgp16}) into the 
GPE equations (\ref{kge13})-(\ref{kge14}), and 
separating real and imaginary
parts, we obtain the system of hydrodynamic equations
\begin{eqnarray}
\frac{\partial\rho}{\partial t}+3H\rho+\frac{1}{a}\vec\nabla\cdot (\rho
{\vec v})=\frac{1}{mc^2}\frac{\partial }{\partial t}\left (\rho \frac{\partial
S}{\partial t}\right )
+\frac{3H\rho}{mc^2}\frac{\partial S}{\partial t}+
\frac{4\rho}{mc^4}\frac{\partial\Phi}{\partial t}\left (mc^2-\frac{\partial
S}{\partial t}\right )-\frac{4\Phi}{ac^2}\vec\nabla\cdot (\rho {\vec
v}),\nonumber\\
\label{kge15}
\end{eqnarray}
\begin{eqnarray}
\frac{\partial S}{\partial t}+\frac{(\vec\nabla S)^2}{2 m a^2}=-\frac{\hbar^2}{2
m
c^2}\frac{\frac{\partial^2\sqrt{\rho}}{\partial
t^2}}{\sqrt{\rho}}
+\left (1+\frac{4\Phi}{c^2}\right )\frac{\hbar^2}{2
m a^2}\frac{\Delta\sqrt{\rho}}{\sqrt{\rho}}-\frac{2\Phi}{m c^2
a^2}(\vec\nabla S)^2\nonumber\\
-m\Phi-\left (1+\frac{2\Phi}{c^2}\right ) m h(\rho)
+\frac{1}{2 mc^2}\left (\frac{\partial
S}{\partial t}\right )^2-\left (3H-\frac{4}{c^2}\frac{\partial\Phi}{\partial
t}\right )\frac{\hbar^2}{4 m c^2 \rho}\frac{\partial\rho}{\partial t},
\label{kge18}
\end{eqnarray}
\begin{eqnarray}
\frac{\partial {\vec v}}{\partial t}+H{\vec v}+\frac{1}{a}({\vec v}\cdot
\vec\nabla){\vec v}=-\frac{\hbar^2}{2am^2c^2}\vec\nabla \left
(\frac{\frac{\partial^2\sqrt{\rho}}{\partial t^2}}{\sqrt{\rho}}\right
)
+\frac{\hbar^2}{2m^2a^3}\vec\nabla\left\lbrack \left (1+\frac{4\Phi}{c^2}\right
)\frac{\Delta\sqrt{\rho}}{\sqrt{\rho}}\right\rbrack\nonumber\\
-\frac{1}{a}\vec\nabla\Phi-\frac{1}{\rho a}\vec\nabla p
-\frac{2}{a c^2}\vec\nabla (h\Phi)-\frac{2}{a c^2}\vec\nabla
(\Phi
v^2)+\frac{1}{2am^2c^2}\vec\nabla \left\lbrack\left (\frac{\partial S}{\partial
t}\right )^2\right \rbrack\nonumber\\
-\frac{3\hbar^2}{4 a m^2 c^2} H\vec\nabla \left
(\frac{1}{\rho}\frac{\partial\rho}{\partial t}\right )+\frac{\hbar^2}{a m^2
c^4}\vec\nabla \left (\frac{\partial\Phi}{\partial
t}\frac{1}{\rho}\frac{\partial\rho}{\partial t}\right ),
\label{kge16}
\end{eqnarray}
\begin{eqnarray}
\frac{\Delta\Phi}{4\pi G a^2}=\left (1-\frac{\Phi}{c^2}\right )\rho
+\frac{\hbar^2}{2m^2c^4}\left (1-\frac{2\Phi}{c^2}\right )\left\lbrack
\frac{1}{4\rho}\left (\frac{\partial\rho}{\partial t}\right
)^2+\frac{\rho}{\hbar^2}\left (\frac{\partial S}{\partial t}\right
)^2\right\rbrack\nonumber\\
+\frac{\hbar^2}{2a^2m^2c^2}\left (1+\frac{2\Phi}{c^2}\right )\left\lbrack
\frac{1}{4\rho}(\vec\nabla\rho)^2+\frac{\rho}{\hbar^2}(\vec\nabla
S)^2\right\rbrack
\nonumber\\
+\frac{1}{c^2}V(\rho)-\frac{1}{m c^2}\left
(1-\frac{2\Phi}{c^2}\right )\rho\frac{\partial S}{\partial t}
-\frac{3H^2}{8\pi G}+\frac{3H}{4\pi G c^2}\left
(\frac{\partial\Phi}{\partial
t}+H\Phi\right ),
\label{kge17}
\end{eqnarray}
where $h(\rho)=V'(\rho)$ is a pseudo  enthalpy and $p(\rho)$ is a pseudo
pressure defined by the relation $h'(\rho)=p'(\rho)/\rho$ \cite{prd1}. It is
explicitly given by $p(\rho)=\rho h(\rho)-\int h(\rho)\, d\rho$, i.e.,
\begin{eqnarray}
p(\rho)=\rho V'(\rho)-V(\rho).
\label{kgp20g}
\end{eqnarray}
The pseudo velocity of sound is $c_s^2=p'(\rho)=\rho V''(\rho)$. For the
quartic potential (\ref{tb4}), we have
\begin{eqnarray}
V(\rho)=\frac{2\pi a_s\hbar^2}{m^3}\rho^2, \qquad h(\rho)=\frac{4\pi
a_s\hbar^2}{m^3}\rho,\qquad p(\rho)=\frac{2\pi a_s\hbar^2}{m^3}\rho^2,\qquad 
c_s^2=\frac{4\pi a_s\hbar^2}{m^3}\rho.
\label{kgp20}
\end{eqnarray}
The  pseudo pressure is
given by a polytropic  equation of state of index $\gamma=2$ which is
quadratic. We
note that this equation of state coincides with the equation
of state of a nonrelativistic self-interacting BEC \cite{revuebec}.
This coincidence is not obvious because Eqs. (\ref{kge15})-(\ref{kge17}) are
valid in the relativistic regime. The
interpretation of this equation of state is, however, not direct because $\rho$
and $p$ are a pseudo density and a pseudo pressure that coincide with the real
density
and the real pressure of a BEC only in the nonrelativistic limit $c\rightarrow
+\infty$.

The energy density and the
pressure can be written in terms of hydrodynamic
variables as
\begin{eqnarray}
\epsilon=\frac{\hbar^2}{2m^2c^2}\left (1-\frac{2\Phi}{c^2}\right )\left\lbrack
\frac{1}{4\rho}\left (\frac{\partial\rho}{\partial t}\right
)^2+\frac{\rho}{\hbar^2}\left (\frac{\partial S}{\partial t}\right
)^2\right\rbrack
+\frac{\hbar^2}{2a^2m^2}\left (1+\frac{2\Phi}{c^2}\right )\left\lbrack
\frac{1}{4\rho}(\vec\nabla\rho)^2+\frac{\rho}{\hbar^2}(\vec\nabla
S)^2\right\rbrack\nonumber\\
-\left (1-\frac{2\Phi}{c^2}\right )\frac{\rho}{m}\frac{\partial S}{\partial
t}+\frac{1}{2}\left (1-\frac{2\Phi}{c^2}\right )\rho
c^2+\frac{1}{2}\rho c^2+V(\rho),\qquad 
\label{kgp22}
\end{eqnarray}
\begin{eqnarray}
P=\frac{\hbar^2}{2m^2c^2}\left (1-\frac{2\Phi}{c^2}\right )\left\lbrack
\frac{1}{4\rho}\left (\frac{\partial\rho}{\partial t}\right
)^2+\frac{\rho}{\hbar^2}\left (\frac{\partial S}{\partial t}\right
)^2\right\rbrack
-\frac{\hbar^2}{6a^2m^2}\left (1+\frac{2\Phi}{c^2}\right )\left\lbrack
\frac{1}{4\rho}(\vec\nabla\rho)^2+\frac{\rho}{\hbar^2}(\vec\nabla
S)^2\right\rbrack\nonumber\\
-\left (1-\frac{2\Phi}{c^2}\right )\frac{\rho}{m}\frac{\partial S}{\partial
t}+\frac{1}{2}\left (1-\frac{2\Phi}{c^2}\right )\rho c^2-\frac{1}{2}\rho
c^2-V(\rho).\qquad 
\label{kgp23}
\end{eqnarray}
We note that, in general, the pressure $P$ defined by Eq. (\ref{kgp23}) is
different from the
pressure $p$ defined by Eq. (\ref{kgp20g}).\footnote{In our
formalism, $p$ represents the pressure
arising from the self-interaction of the bosons (scattering) while $P$ is the
pressure of the SF defined by analogy with the pressure of an ideal
fluid whose energy momentum tensor writes
$T_{\mu\nu}=(P+\epsilon)u_{\mu}u_{\nu}/c^2-Pg_{\mu\nu}$.} However,
they coincide for a homogeneous SF in the regime where the SF
oscillations are faster than the Hubble expansion
\cite{abrilph1,abrilph2}.

The hydrodynamic equations  (\ref{kge15})-(\ref{kge17}) have a clear physical
interpretation. Eq. (\ref{kge15}), corresponding to the imaginary
part of the GPE equations, is the continuity equation. We note that
$\int \rho\, d^3x$ is not conserved in the relativistic regime. However,  Eq.
(\ref{kge15}) is consistent with the
conservation of the charge of a SF (see section
\ref{sec_b} and \ref{sec_c}). Eq.
(\ref{kge18}),
corresponding to the real part of the GPE equations,
is the Bernoulli or Hamilton-Jacobi equation. Eq. (\ref{kge16}), obtained by
taking the gradient of  Eq. (\ref{kge18}), is the momentum equation. Eq.
(\ref{kge17}) is the Einstein equation. We stress
that the hydrodynamic equations  (\ref{kge15})-(\ref{kge17}) are equivalent to
the GPE equations (\ref{kge13})-(\ref{kge14}) which are themselves equivalent
to the 
KGE equations (\ref{kge4}) and (\ref{kge12}). In the nonrelativistic limit
$c\rightarrow +\infty$, we recover the quantum Euler-Poisson equations
\cite{chavaniscosmo}:
\begin{eqnarray}
\frac{\partial\rho}{\partial t}+3H\rho+\frac{1}{a}\vec\nabla\cdot (\rho
{\vec v})=0,
\label{nr3}
\end{eqnarray}
\begin{eqnarray}
\frac{\partial S}{\partial t}+\frac{(\vec\nabla S)^2}{2 m
a^2}=\frac{\hbar^2}{2
m a^2}\frac{\Delta\sqrt{\rho}}{\sqrt{\rho}}
-m\Phi-mh(\rho),
\label{nr6}
\end{eqnarray}
\begin{eqnarray}
\frac{\partial {\vec v}}{\partial t}+H{\vec v}+\frac{1}{a}({\vec v}\cdot
\vec\nabla){\vec v}=
\frac{\hbar^2}{2m^2a^3}\vec\nabla\left( \frac{\Delta\sqrt{\rho}}{\sqrt{
\rho}} \right )-\frac{1}{a}
\vec\nabla\Phi-\frac{1}{\rho a}\vec\nabla p,
\label{nr4}
\end{eqnarray}
\begin{eqnarray}
\frac{\Delta\Phi}{4\pi G a^2}=\rho-\frac{3H^2}{8\pi
G}.
\label{nr5}
\end{eqnarray}
These equations can also be written in terms of the density contrast
$\delta=(\rho-\rho_b)/\rho_b$ \cite{chavaniscosmo}.

\section{Cosmological evolution of a spatially homogeneous scalar field}
\label{sec_b}

We consider the evolution of a universe induced solely by a spatially
homogeneous
SF. In the comoving frame, we have $\rho(\vec x,t)=\rho_b(t)$,
$\vec v_b(\vec x,t)=\vec 0$, $\Phi_b(\vec x,t)=0$, and
$S_b(\vec x,t)=S_b(t)$. We introduce the notation $E(t)=-dS_b/dt$ which can
be considered as the time-dependent energy of the spatially homogeneous SF in
the comoving frame. The pseudo wave function of the SF is $\psi_b(\vec
x,t)=\psi_b(t)=\sqrt{\rho_b(t)}e^{-(i/\hbar)\int E(t)\, dt}$. Using Eq.
(\ref{kgp12}),
we have $\varphi_b(\vec
x,t)=\varphi_b(t)=\frac{\hbar}{m}\sqrt{\rho_b(t)}e^{-(i/\hbar)\lbrack m c^2
t + \int E(t)\, dt\rbrack}$ so the total energy of the SF, including its
rest mass, is $E_{\rm tot}(t)=E(t)+mc^2$.

For a spatially homogeneous SF, the hydrodynamic equations
(\ref{kge15})-(\ref{kge17})  reduce to
\begin{eqnarray}
\frac{d\rho_b}{dt}+3H\rho_b=-\frac{1}{mc^2}\frac{d}{dt}\left
(\rho_b E\right )-\frac{3H\rho_b} {
mc^2}E,
\label{b1}
\end{eqnarray}
\begin{eqnarray}
\left (\frac{E}{2 mc^2}+1\right )E=\frac{\hbar^2}{2 m
c^2}\frac{\frac{d^2\sqrt{\rho_b}}{d
t^2}}{\sqrt{\rho_b}}
+mh(\rho_b)
+\frac{3H\hbar^2}{4 m c^2
\rho_b}\frac{d\rho_b}{dt},
\label{b2}
\end{eqnarray}
\begin{eqnarray}
\frac{3H^2}{8\pi G}=\rho_b
+\frac{\hbar^2}{2m^2c^4}\left\lbrack
\frac{1}{4\rho_b}\left (\frac{d\rho_b}{d t}\right
)^2+\frac{\rho_b}{\hbar^2}E^2\right\rbrack
+\frac{1}{c^2}V(\rho_b)+\frac{E}{m
c^2}\rho_b.
\label{b3}
\end{eqnarray}

In terms of the total energy $E_{\rm tot}(t)=E(t)+mc^2$, the equation of
continuity (\ref{b1}) becomes
\begin{eqnarray}
\frac{1}{\rho_b}\frac{d\rho_b}{dt}+\frac{3}{a}\frac{da}{dt}+\frac{1}{E_{\rm
tot}} \frac{dE_{\rm tot}}{dt}=0.
\label{b4}
\end{eqnarray}
It can be rewritten as a conservation law:
\begin{eqnarray}
\frac{d}{dt}(E_{\rm tot}\, \rho_b
a^3)=0. 
\label{b5}
\end{eqnarray}
Therefore, the total energy is exactly given by
\begin{eqnarray}
\frac{E_{\rm tot}}{mc^2}=\frac{Qm}{\rho_b a^3},
\label{b6}
\end{eqnarray}
where $Q$ is a constant. This conservation law was found by Gu and Hwang
\cite{gh} directly from the KG equation. It can be shown
that $Q= \int J^0\, d^3x$ represents the
conserved charge
density of the complex SF (see \ref{sec_c}). The energy density
and the
pressure of a homogeneous SF are
\begin{eqnarray}
\epsilon_b=\frac{\hbar^2}{8m^2c^2}\frac{1}{\rho_b}\left
(\frac{d\rho_b}{dt}\right
)^2+\frac{\rho_b}{m}E\left (1+\frac{E}{2mc^2}\right
)+\rho_b c^2+V(\rho_b),
\label{kgp22h}
\end{eqnarray}
\begin{eqnarray}
P_b=\frac{\hbar^2}{8m^2c^2}\frac{1}{\rho_b}\left
(\frac{d\rho_b}{dt}\right
)^2+\frac{\rho_b}{m}E\left (1+\frac{E}{2mc^2}\right
)-V(\rho_b).
\label{kgp23h}
\end{eqnarray}
Equations (\ref{b2}), (\ref{b3}) and (\ref{b6}) determine the complete
evolution of a universe induced by a spatially homogeneous SF. Working
directly on
the homogeneous KG equation with a quartic self-interaction potential, Li {\it
et al.} \cite{shapiro} have shown that
a universe filled with a relativistic complex SF first undergoes an intrinsic
stiff matter era, followed by a radiation era due to its self-interaction,
before finally entering in the matter era. The stiff matter era occurs when the 
SF oscillations are slower than the Hubble expansion while the radiation and
matter eras occur when the SF oscillations are faster than the Hubble
expansion. These different regimes can be recovered from the
hydrodynamic equations (\ref{b1})-(\ref{b3}) \cite{abrilph1}. 

In the nonrelativistic limit $c\rightarrow +\infty$, Eqs. (\ref{b1})-(\ref{b3})
reduce to
\begin{eqnarray}
\frac{d\rho_b}{dt}+3H\rho_b=0,\qquad E=mh(\rho_b),\qquad
\frac{3H^2}{8\pi G}=\rho_b.
\label{sun1}
\end{eqnarray}
We find that $\rho_b\propto a^{-3}$, $a\propto t^{2/3}$ and $\rho_b=1/(6\pi
Gt^2)$ (Einstein-de Sitter solution), so the homogeneous SF/BEC behaves as CDM.
For the quartic potential (\ref{tb4}), we
have $E(t)={4\pi a_s
\hbar^2\rho_b}/{m^2}=2a_s\hbar^2/3Gm^2t^2$ and $S_b(t)=2a_s\hbar^2/3Gm^2t+C$.

\section{Generalized Klein-Gordon-Poisson equations}
\label{sec_kgp}

In this section, we consider a simplified model in which we introduce the
gravitational potential $\Phi(\vec x,t)$ in the ordinary KG
equation by hand, as an external potential, and assume that this potential
is produced
by the SF itself via a generalized Poisson equation in which the source is the
energy density $\epsilon$. This leads to
the generalized KGP equations. We then show that these equations can be
rigorously justified from the KGE
equations in the limit $\Phi/c^2\rightarrow 0$. However, this
simplified model is not sufficient to study the evolution of the perturbations
in the linear relativistic regime since it precisely neglects terms of order
$\Phi/c^2$.

We consider the FLRW metric that describes an isotropic and
homogeneous expanding background. The line element in the comoving frame is
\begin{equation}
ds^2=g_{\mu\nu}dx^{\mu}dx^{\nu}=c^2dt^2-a(t)^2\delta_{ij}dx^idx^j.
\label{kgp1}
\end{equation}
For this metric, the d'Alembertian operator
(\ref{tb7}) writes
\begin{equation}
\Box=\frac{1}{c^2}\frac{\partial^2}{\partial
t^2}+\frac{3H}{c^2}\frac{\partial}{\partial t}-\frac{1}{a^2}\Delta.
\label{kgp2}
\end{equation}
In order to take the self-gravity of the SF into account, we
introduce a Lagrangian of interaction that couples the gravitational potential
$\Phi(\vec x,t)$ to the scalar
field $\varphi(\vec x,t)$ according to
\begin{equation}
\mathcal{L}_{\rm int}=-\frac{m^2}{\hbar^2}\Phi|\varphi|^2.
\label{kgp3}
\end{equation} 
The total Lagrangian of the system (SF $+$ gravity) is
given by $\mathcal{L}=\mathcal{L}_\varphi+\mathcal{L}_{\rm int}$. The equation
of motion resulting from the stationarity of the
total action $S=S_{\varphi}+S_{\rm int}$, obtained by writing $\delta
S=0$, is the KG equation
\begin{equation}
\Box\varphi+\frac{m^2c^2}{\hbar^2}\varphi+2V(|\varphi|^2),_{\varphi^*}+\frac{
2m^2 } { \hbar^2 } \Phi\varphi=0 ,
\label{kgp5}
\end{equation}
where the d'Alembertian operator is given by Eq. (\ref{kgp2}) and the
gravitational
potential  $\Phi(\vec x,t)$ acts here as an external potential.
The energy density
and the pressure, defined from the diagonal part of the
energy-momentum tensor (\ref{tb12}), are given by
\begin{equation}
\epsilon=\frac{1}{2c^2}\left |\frac{\partial\varphi}{\partial
t}\right|^2+\frac{1}{2a^2}|\vec\nabla\varphi|^2+\frac{m^2c^2}{2\hbar^2}
|\varphi|^2+V(|\varphi|^2),
\label{kgp7}
\end{equation}
\begin{equation}
P=\frac{1}{2c^2}\left |\frac{\partial\varphi}{\partial
t}\right|^2-\frac{1}{6a^2}|\vec\nabla\varphi|^2-\frac{m^2c^2}{2\hbar^2}
|\varphi|^2-V(|\varphi|^2).
\label{kgp8}
\end{equation}

Eq. (\ref{kgp5}) is the ordinary KG equation for a SF in an
external potential $\Phi(\vec x,t)$ in an expanding background. We now state
that $\Phi(\vec x,t)$ is actually the gravitational potential produced by the
SF itself. We phenomenologically assume that the gravitational
potential is determined by a generalized Poisson equation of the
form
\begin{eqnarray}
\frac{\Delta\Phi}{4\pi
Ga^2}=\frac{1}{c^2}(\epsilon-\epsilon_b)
\label{kgp9}
\end{eqnarray}
in which the source of the gravitational potential is the energy density
$\epsilon$ of the SF (more precisely, its deviation from the
homogeneous background density $\epsilon_b(t)$). Using Eq. (\ref{kgp7}) for the
energy
density of a SF, and recalling the Friedmann equation (\ref{kge12b}),
the generalized Poisson equation can be written
as
\begin{eqnarray}
\frac{\Delta\Phi}{4\pi
Ga^2}=\frac{1}{2c^4}\left |\frac{\partial\varphi}{\partial
t}\right |^2+\frac{1}{2a^2c^2}|\vec\nabla\varphi|^2+\frac{m^2}{2\hbar^2}
|\varphi|^2
+\frac{1}{c^2}V(|\varphi|^2)-\frac{3H^2}{8\pi
G}.
\label{kgp11}
\end{eqnarray}
Eqs. (\ref{kgp5}) and (\ref{kgp11}) form the generalized KGP equations. They
have been introduced in an {\it ad hoc} manner but they
can be rigorously justified from the  KGE equations 
(\ref{kge4}) and (\ref{kge12}) in the limit $\Phi/c^2\rightarrow 0$
(which, of course, is different from the nonrelativistic limit $c\rightarrow
+\infty$). We see that the gravitational potential $\Phi$ appears in the KG
equation (\ref{kgp5}) due to the cancelation of $c^2$ in the product
$\Phi/c^2\times c^2$ in Eq. (\ref{kge4}). Therefore, we do not have to
introduce $\Phi$ by hand: the generalized KGP equations can be
obtained from the KGE equations  by simply neglecting terms of order $\Phi/c^2$
in these equations. Similarly, the equations related to the generalized KGP
equations can be
obtained from the ones related to the KGE equations by neglecting terms of
order $\Phi/c^2$. For example, the generalized GPP equations write
\begin{eqnarray}
i\hbar\frac{\partial\psi}{\partial t}-\frac{\hbar^2}{2m
c^2}\frac{\partial^2\psi}{\partial t^2}-\frac{3}{2}H\frac{\hbar^2}{m
c^2}\frac{\partial\psi}{\partial t}
+\frac{\hbar^2}{2 m a^2}\Delta\psi-m\Phi \psi
-m \frac{dV}{d|\psi|^2}\psi+\frac{3}{2}i\hbar
H\psi=0,
\label{kge13p}
\end{eqnarray}
\begin{eqnarray}
\frac{\Delta\Phi}{4\pi G a^2}=|\psi|^2
+\frac{\hbar^2}{2m^2c^4}\left
|\frac{\partial\psi}{\partial t}\right
|^2+\frac{\hbar^2}{2a^2m^2c^2}|\vec\nabla\psi|^2
+\frac{1}{c^2}V(|\psi|^2)-\frac{\hbar}{m c^2}{\rm Im} \left
(\frac{\partial\psi}{\partial
t}\psi^*\right ) -\frac{3H^2}{8\pi G}.
\label{kge14p}
\end{eqnarray}
The energy density and the pressure are given by
\begin{eqnarray}
\epsilon=\frac{\hbar^2}{2m^2c^2}\left
|\frac{\partial\psi}{\partial t}\right |^2
-\frac{\hbar}{m}{\rm Im} \left
(\frac{\partial\psi}{\partial
t}\psi^*\right )
+\frac{\hbar^2}{2a^2m^2}|\vec\nabla\psi|^2
+c^2|\psi|^2+V(|\psi|^2),
\label{kgp14bsimp}
\end{eqnarray}
\begin{eqnarray}
P=\frac{\hbar^2}{2m^2c^2}\left
|\frac{\partial\psi}{\partial t}\right |^2
-\frac{\hbar}{m}{\rm Im} \left
(\frac{\partial\psi}{\partial
t}\psi^*\right )
-\frac{\hbar^2}{6a^2m^2}|\vec\nabla\psi|^2-V(|\psi|^2).
\label{kgp14csimp}
\end{eqnarray}

The corresponding hydrodynamic equations write
\begin{eqnarray}
\frac{\partial\rho}{\partial t}+3H\rho+\frac{1}{a}\vec\nabla\cdot (\rho
{\vec v})=\frac{1}{mc^2}\frac{\partial }{\partial t}\left (\rho \frac{\partial
S}{\partial t}\right )
+\frac{3H\rho}{mc^2}\frac{\partial S}{\partial t},
\label{kge15p}
\end{eqnarray}
\begin{equation}
\frac{\partial S}{\partial t}+\frac{(\vec\nabla S)^2}{2 m a^2}=-\frac{\hbar^2}{2
m
c^2}\frac{\frac{\partial^2\sqrt{\rho}}{\partial
t^2}}{\sqrt{\rho}}
+\frac{\hbar^2}{2
m a^2}\frac{\Delta\sqrt{\rho}}{\sqrt{\rho}}
-m\Phi- m h(\rho)
+\frac{1}{2 mc^2}\left (\frac{\partial
S}{\partial t}\right )^2-\frac{3H\hbar^2}{4 m c^2
\rho}\frac{\partial\rho}{\partial t},
\label{kge18p}
\end{equation}
\begin{eqnarray}
\frac{\partial {\vec v}}{\partial t}+H{\vec v}+\frac{1}{a}({\vec v}\cdot
\vec\nabla){\vec v}=-\frac{\hbar^2}{2am^2c^2}\vec\nabla \left
(\frac{\frac{\partial^2\sqrt{\rho}}{\partial t^2}}{\sqrt{\rho}}\right
)
+\frac{\hbar^2}{2m^2a^3}\vec\nabla\left ( \frac{\Delta\sqrt{\rho}}{\sqrt{
\rho}} \right )-\frac{1}{a}
\vec\nabla\Phi-\frac{
1}{\rho a}\vec\nabla p\nonumber\\
+\frac{1}{2am^2c^2}\vec\nabla \left\lbrack\left
(\frac{\partial S}{\partial
t}\right )^2\right \rbrack
-\frac{3\hbar^2}{4 a m^2 c^2} H\vec\nabla \left
(\frac{1}{\rho}\frac{\partial\rho}{\partial t}\right
),
\label{kge16p}
\end{eqnarray}
\begin{eqnarray}
\frac{\Delta\Phi}{4\pi G a^2}=\rho
+\frac{\hbar^2}{2m^2c^4}\left\lbrack
\frac{1}{4\rho}\left (\frac{\partial\rho}{\partial t}\right
)^2+\frac{\rho}{\hbar^2}\left (\frac{\partial S}{\partial t}\right
)^2\right\rbrack
+\frac{\hbar^2}{2a^2m^2c^2}\left\lbrack
\frac{1}{4\rho}(\vec\nabla\rho)^2+\frac{\rho}{\hbar^2}(\vec\nabla
S)^2\right\rbrack
\nonumber\\
+\frac{1}{c^2}V(\rho)-\frac{1}{m c^2}\rho\frac{\partial S}{\partial t}
-\frac{3H^2}{8\pi G}.
\label{kge17p}
\end{eqnarray}
The energy density and the
pressure can be written in terms of hydrodynamic
variables as
\begin{equation}
\epsilon=\frac{\hbar^2}{2m^2c^2}\left\lbrack
\frac{1}{4\rho}\left (\frac{\partial\rho}{\partial t}\right
)^2+\frac{\rho}{\hbar^2}\left (\frac{\partial S}{\partial t}\right
)^2\right\rbrack
+\frac{\hbar^2}{2a^2m^2}\left\lbrack
\frac{1}{4\rho}(\vec\nabla\rho)^2+\frac{\rho}{\hbar^2}(\vec\nabla
S)^2\right\rbrack
-\frac{\rho}{m}\frac{\partial S}{\partial
t}+\rho
c^2+V(\rho),
\label{kgp22simple}
\end{equation}
\begin{equation}
P=\frac{\hbar^2}{2m^2c^2}\left\lbrack
\frac{1}{4\rho}\left (\frac{\partial\rho}{\partial t}\right
)^2+\frac{\rho}{\hbar^2}\left (\frac{\partial S}{\partial t}\right
)^2\right\rbrack
-\frac{\hbar^2}{6a^2m^2}\left\lbrack
\frac{1}{4\rho}(\vec\nabla\rho)^2+\frac{\rho}{\hbar^2}(\vec\nabla
S)^2\right\rbrack
-\frac{\rho}{m}\frac{\partial S}{\partial
t}-V(\rho).
\label{kgp23simple}
\end{equation}

This model correctly describes the homogeneous background for which $\Phi=0$ but
it  is not sufficient to describe the evolution of the perturbations in the
linear regime because we must precisely take into account the terms
of order
$\Phi/c^2$ in this regime (except, of course, in the nonrelativistic limit
$c\rightarrow +\infty$). Therefore, the
use of the KGE equations is mandatory to study the evolution of the
perturbations in the relativistic regime.

{\it Remark:} We could also assume that the
gravitational potential is determined by a Poisson equation 
of the form
\begin{eqnarray}
\frac{\Delta\Phi}{4\pi
Ga^2}=\rho
\label{kgp9w}
\end{eqnarray}
in which the source of the gravitational potential is the pseudo rest-mass
density $\rho=|\psi|^2$ of the SF.  Eqs. (\ref{kgp5}) and (\ref{kgp9w}) form the
KGP equations. This approximation has been considered in \cite{abrilMNRAS}.
However, there is an inconsistency in coupling the relativistic KG equation
(\ref{kgp5}) to the classical Poisson equation (\ref{kgp9w}).

\section{Conclusion}
\label{sec_conclusion}

We have developed  a formalism based on a relativistic
SF described by the KGE equations in the weak field limit. We have transformed
these equations into equivalent hydrodynamic
equations. These equations are arguably more tractable than the KGE equations
themselves. In the nonrelativistic limit, they reduce to the  hydrodynamic
equations directly obtained from the GPP equations \cite{chavaniscosmo}.
Therefore, our
formalism clarifies the
connection between the relativistic and nonrelativistic treatments. We note
that, in the relativistic regime, the hydrodynamic variables $\psi$, $\rho$,
$S$, $\vec v$, $p$,... that we have introduced do not have a direct physical
interpretation. It is only in the nonrelativistic limit $c\rightarrow
+\infty$ that they coincide with
the wave function, rest-mass density, action, velocity, and pressure. However,
these variables are perfectly well-defined mathematically from the SF 
$\varphi$ in any regime, and they are totally legitimate. Furthermore, in terms
of these variables, the relativistic hydrodynamic equations take a relatively
simple form that provides a natural generalization of the nonrelativistic
hydrodynamic equations.

The complete study of these relativistic hydrodynamic equations is of
considerable interest but it is, of course, of great complexity. In our research
papers \cite{abrilph1,abrilph2}, we have started their study in simple cases. 
We
have checked that the hydrodynamic equations of the SFDM
model reproduce the evolution of the homogeneous background obtained previously
by Li {\it et al.} \cite{shapiro} directly from the KGE equations: a stiff
matter era, followed by a radiation era (for a self-interacting SF), and a
matter era. We have also started to study the evolution of the perturbations
in the linear regime in a static and in an expanding universe. We
have shown analytically and numerically that perturbations whose wavelength is
below the Jeans length
oscillate in time while perturbations whose wavelength is above the Jeans
length grow linearly with the scale factor as in the CDM
model. The growth of perturbations in the SF model is substantially
faster than in the CDM model. We have
also shown that general relativity attenuates or even prevents the growth of
perturbations at very large scales, close to the horizon (Hubble length). 

For physically  relevant wavelengths, the nonrelativistic limit of our formalism
is sufficient to describe the evolution of the perturbations in the matter era.
However, even if relativistic corrections are weak in the matter era, we may
wonder whether their effect could be detected in an era of precision
cosmology. In particular, it would be interesting to see if
one
can observe differences between the KGE equations considered in these
Proceedings and
the heuristic KGP equations studied  in the past, in which gravity is introduced
by hand in the KG
equations (see section \ref{sec_kgp}).

In future works, it will be important to consider
the nonlinear
regime where structure formation actually
occurs. In
general, this problem must
be addressed numerically. The hydrodynamic equations
derived in these Proceedings may be very helpful because they may be easier
to solve than the KGE equations. Therefore, numerical
simulations
using fluid dynamics should be developed in the future. As a first step,
relativistic effects could be neglected and the nonrelativistic
equations of Ref.
\cite{chavaniscosmo} could be used. These
equations are similar to the
hydrodynamic equations of CDM
except that they include a quantum potential (Heisenberg) and a pressure term
(scattering) that avoid singularities at small scales \cite{chavaniscosmo}. In
this respect, it may be recalled that the SP equations
were introduced early by Widrow and Kaiser \cite{widrow} as a
procedure of small-scale regularization (a sort of mathematical trick) to
prevent singularities in collisionless simulations of {\it
classical} particles. In their approach, $\hbar$ is not the Planck
constant, but rather an ajustable parameter that controls the spatial
resolution. Their procedure may find a
physical justification if DM is made of self-gravitating BECs \cite{chavkpz}.

Our relativistic formalism may have applications for other self-gravitating
systems described by SFs or BECs besides DM. We can mention, for
example, the case of boson
stars \cite{kaup,rb,colpi} and the
case of
microscopic quantum black holes made of BECs of gravitons stuck at a
quantum critical point \cite{dvali,casadio,bookspringer}. It has
also
been proposed \cite{chavharko} that, because of their superfluid core, neutron
stars could be BEC
stars. Indeed, the neutrons (fermions) could form Cooper pairs and behave as
bosons of mass $2m_n$. This idea may solve certain issues regarding the maximum
mass of neutron stars. Finally, we may mention analog 
models of gravity in which BECs described by the GP
equation or by the KG equation are used to simulate
results of classical and quantum field theory in curved spacetime
\cite{bl2}.

\ack A. S. acknowledges CONACyT for the postdoctoral
grant received.

\appendix

\section{The value of $A$}
\label{sec_cons}

The GP equation is obtained from the KG equation by means of the transformation
\begin{eqnarray}
\varphi=A e^{-i m c^2
t/\hbar}\psi.
\label{k1}
\end{eqnarray}
The constant $A$ can be computed as follows. Substituting Eq. (\ref{k1})
into Eq. (\ref{kge5}), we find that the energy density of the SF is
given by
\begin{eqnarray}
\frac{\epsilon}{c^2}=\frac{T_0^0}{c^2}=\frac{1}{2}\left
(1-\frac{2\Phi}{c^2}\right
)\frac{m^2}{\hbar^2}A^2|\psi|^2+\frac{m^2}{2\hbar^2}A^2
|\psi|^2
+\frac{A^2}{2c^4}\left (1-\frac{2\Phi}{c^2}\right )\left
|\frac{\partial\psi}{\partial t}\right |^2\nonumber\\
+\frac{A^2}{2a^2c^2}\left
(1+\frac{2\Phi}{c^2}\right )|\vec\nabla\psi|^2
+\frac{1}{c^2}V(|\psi|^2)-\frac{m A^2}{\hbar
c^2}\left (1-\frac{2\Phi}{c^2}\right ){\rm Im} \left
(\frac{\partial\psi}{\partial t}\psi^*\right ).
\label{k2}
\end{eqnarray}
Taking the nonrelativistic limit $c\rightarrow +\infty$ of this equation, we
obtain
\begin{eqnarray}
\frac{\epsilon}{c^2}\rightarrow
\frac{m^2A^2}{\hbar^2}|\psi|^2=\frac{m^2 A^2}{\hbar^2}\rho,
\label{k3}
\end{eqnarray}
where $\rho=|\psi|^2$ is the rest-mass density. Since $\epsilon\sim\rho c^2$ in
the nonrelativistic limit $c\rightarrow +\infty$, we find
\begin{eqnarray}
A=\frac{\hbar}{m}.
\label{k5}
\end{eqnarray}

\section{Some comments about the Klein-Gordon equation}
\label{sec_c}

In this Appendix, we recall the difficulties associated with the
interpretation of the KG equation. We also clarify the relation between the
hydrodynamic representation of the KG equation and the conservation of the
charge.

The fundamental equation of nonrelativistic quantum mechanics is the
Schr\"odinger \cite{schrodinger1} equation
\begin{eqnarray}
i\hbar\frac{\partial\psi}{\partial t}=-\frac{\hbar^2}{2m}\Delta\psi.
\label{c1}
\end{eqnarray}
If we write the wave function under the form
$\psi(\vec{x},t)=\sqrt{\rho(\vec{x},t)} e^{iS(\vec{x},t)/\hbar}$ and define
the density $\rho$ and the current $\vec{J}$ by
\begin{eqnarray}
\rho=|\psi|^2,
\label{c2}
\end{eqnarray}
\begin{eqnarray}
\vec{J}=\rho\frac{\vec\nabla S}{m}=\frac{\hbar}{2 i
m}(\psi^*\vec\nabla\psi-\psi\vec\nabla\psi^*),
\label{c3}
\end{eqnarray}
where we have used $S=(\hbar/2i)\ln(\psi/\psi^*)$, we obtain the continuity
equation
\begin{eqnarray}
\frac{\partial\rho}{\partial t}+\nabla\cdot \vec{J}=0.
\label{c4}
\end{eqnarray}
This equation shows that the integral of the density $\int \rho d^3{x}$ is
conserved. Furthermore, by definition, the density is positive:
$\rho(\vec{x},t)\ge 0$. Therefore, $\rho(\vec{x},t)$ can be interpreted as a
probability density. 

In the Madelung hydrodynamic representation of the Schr\"odinger equation, we
define the density $\rho$ and the velocity $\vec{v}$ by
\begin{eqnarray}
\rho=|\psi|^2,\qquad \vec v=\frac{\vec \nabla S}{m}.
\label{c5}
\end{eqnarray}
These variables satisfy the continuity equation
\begin{eqnarray}
\frac{\partial\rho}{\partial t}+\nabla\cdot (\rho \vec{v})=0.
\label{c6}
\end{eqnarray}
Since $\vec{J}=\rho\vec{v}$ according to Eqs. (\ref{c3}) and (\ref{c5}), we
immediately see the  equivalence between Eqs. (\ref{c4}) and (\ref{c6}).

We now consider the KG \cite{klein1,gordon} equation
\begin{eqnarray}
\frac{1}{c^2}\frac{\partial^2\varphi}{\partial
t^2}-\Delta\varphi+\frac{m^2c^2}{\hbar^2}\varphi=0
\label{c7}
\end{eqnarray}
which was initially proposed as a relativistic extension of the Schr\"odinger
equation.\footnote{The KG
equation was actually discovered by Schr\"odinger
before he found the equation that now bears his name \cite{zee}. The
KG equation was also obtained by Fock \cite{fock}, de Donder and van
den Dungen \cite{donder}, and Kudar \cite{kudar}.} If we write the SF under the
form $\varphi(\vec{x},t)=\sqrt{R(\vec{x},t)} e^{i\sigma(\vec{x},t)/\hbar}$ and
introduce the quadricurrent 
$J^{\mu}=-R\partial^{\mu}\sigma/m=-(\hbar/2im)(\varphi^*\partial^{\mu}
\varphi-\varphi\partial^ { \mu }
\varphi^*)$, we obtain the continuity equation  $\partial_{\mu}J^{\mu}=0$.
The quadricurrent can be written as $J^{\mu}=(J^0,\vec{J})$ with 
\begin{eqnarray}
J^0=-R\frac{\frac{1}{c}\frac{\partial\sigma}{\partial t}}{m}=-\frac{\hbar}{2 i
mc}\left (\varphi^*\frac{\partial\varphi}{\partial
t}-\varphi\frac{\partial\varphi^*}{\partial t}\right ),
\label{c8}
\end{eqnarray}
\begin{eqnarray}
\vec{J}=R\frac{\vec\nabla \sigma}{m}=\frac{\hbar}{2 i
m}(\varphi^*\vec\nabla\varphi-\varphi\vec\nabla\varphi^*).
\label{c9}
\end{eqnarray}
If we introduce the KG density $\rho_{\rm KG}=J^0/c$, we can write the
continuity equation as
\begin{eqnarray}
\frac{\partial\rho_{\rm KG}}{\partial t}+\nabla\cdot \vec{J}=0.
\label{c10}
\end{eqnarray}
This equation  shows that the integral of the KG density $\int \rho_{\rm KG}
d^3{x}$ is conserved. However, the KG density  $\rho_{\rm
KG}(\vec{x},t)$ is {\it not} definite positive so it cannot be interpreted as a
density probability. Another difficulty with the KG equation  is that
it allows negative kinetic energies as solution. Indeed, decomposing Eq. 
(\ref{c7}) into plane waves, we obtain two solutions of the form
$\varphi_{\pm}(\vec{x},t)=A_{\pm}
e^{i(\vec{p}\cdot \vec{x}-E_{\pm}t)/\hbar}$ with $E_{\pm}=\pm
\sqrt{p^2c^2+m^2c^4}$.

The original difficulties encountered with the KG equation had interesting
historical developments. Dirac \cite{dirac1} proposed another relativistic
extension of the Schr\"odinger equation. The
Dirac equation describes spin-$1/2$ massive particles (fermions) such as
electrons and quarks. In Dirac's theory, the probability density is positive but
negative
energies
are allowed. Dirac solved the problem of negative energies via the ``hole''
theory \cite{dirac2}. This leads to the concept of antiparticles that are
related to negative energy eigenstates. Antimatter was unsuspected before
Dirac's work.
The first antiparticle, the positron, was experimentally discovered by Anderson
\cite{anderson} in 1932.
On the other hand,
Pauli and Weisskopf \cite{pw} proposed a new interpretation of the KG equation.
They interpreted the KG density  $\rho_{\rm KG}=J^0/c$ as a charge density which
can be of
arbitrary sign. If we define the charge density and
the charge current by  $\rho_e=meJ^0/c\hbar^2$ and
$\vec{J}_e=me\vec{J}/\hbar^2$, where $e$ is an
elementary charge, we can rewrite the continuity equation (\ref{c10}) as
\begin{eqnarray}
\frac{\partial\rho_e}{\partial t}+\nabla\cdot \vec{J}_e=0.
\label{c11}
\end{eqnarray}
This equation expresses the conservation of the total charge $Q_e=\int \rho_e \,
d^3{x}$. Although the KG equation is not a successful relativistic
generalization of the Schr\"odinger equation (initially introduced to
describe the energy spectrum of the electron), this equation was resurrected in
the context of quantum
field theory where it was shown to describe spin-$0$
particles (bosons) such as $\pi$-mesons, pions, or the Higgs boson. In this
interpretation, since $\rho_e^{(\pm)}=\pm
({e|E_{\pm}|}/{\hbar^2c^2})|\varphi_{\pm}|^2$, $\varphi_+$ specifies
particles with charge $+e$ and energy $+E$
while $\varphi_-$ specifies antiparticles with the same mass but with charge
$-e$ and energy $-E$. For a real SF the charge is zero since  $\rho_e=0$ and
$\vec J_e=\vec 0$ according to Eqs. (\ref{c8}) and (\ref{c9}).

In the hydrodynamic representation of the KG equation, we write
\begin{eqnarray}
\varphi(\vec{x},t)=\frac{\hbar}{m}\sqrt{\rho(\vec{x},t)}
e^{i\left\lbrack S(\vec{x},t)-mc^2 t\right\rbrack/\hbar}
\label{c12}
\end{eqnarray}
and define the density $\rho$ and the velocity $\vec{v}$ by 
\begin{eqnarray}
\rho=\frac{m^2}{\hbar^2}|\varphi|^2,\qquad \vec v=\frac{\vec \nabla S}{m}.
\label{c13}
\end{eqnarray}
These variables satisfy the continuity equation
\begin{eqnarray}
\frac{\partial\rho}{\partial t}+\nabla\cdot (\rho
\vec{v})=\frac{1}{mc^2}\frac{\partial}{\partial t}\left (\rho\frac{\partial
S}{\partial t}\right ).
\label{c14}
\end{eqnarray}
The density $\rho(\vec x,t)$ is positive by definition but it is not conserved.
Therefore, $\rho(\vec x,t)$ cannot be interpreted as a density probability.
However, the continuity equation (\ref{c14}) can be rewritten as
\begin{eqnarray}
\frac{\partial}{\partial t}\left (\rho \frac{E_{\rm tot}}{mc^2}\right
)+\nabla\cdot (\rho
\vec{v})=0, 
\label{c15}
\end{eqnarray}
where we have defined $E(\vec x,t)=-\partial S/\partial t$ and $E_{\rm
tot}(\vec x,t)=mc^2+E(\vec x,t)$. This equation implies the conservation
of the integral $\int\rho E_{\rm tot}\,d^3{x}$. To show that this integral
corresponds to the total charge of the SF, we note that $R=(\hbar^2/m^2)\rho$
and $\sigma=S-mc^2 t$. Therefore, 
\begin{eqnarray}
\rho_e=-\frac{e}{c^2\hbar^2}R\frac{\partial\sigma}{\partial
t}=- e\frac{\rho}{m}  \frac{\frac{\partial S}{\partial
t}-mc^2}{mc^2}=e\frac{\rho}{m}  \left (\frac{E}{mc^2}+1\right
)=e\frac{\rho}{m}\frac{E_{\rm tot}}{mc^2},
\label{c16}
\end{eqnarray}
\begin{eqnarray}
\vec{J}_e=\frac{e}{\hbar^2}R\vec\nabla
\sigma=e\frac{\rho}{m}\frac{\vec\nabla
S}{m}=e\frac{\rho}{m}\vec{v}.
\label{c17}
\end{eqnarray}
These relations first establish the equivalence between Eqs. (\ref{c11}),
(\ref{c14}), and (\ref{c15}). Furthermore, they show that the total charge of
the SF can
be written as
\begin{eqnarray}
Q_e=e\int \frac{\rho}{m}\frac{E_{\rm tot}}{mc^2}\,
d^3{x}
\label{c18}
\end{eqnarray}
in agreement with Eq. (\ref{c15}). Eq. (\ref{c18}) is consistent with the
expression (\ref{b6}) of the charge of a spatially homogeneous SF in an
expanding
universe.


\section*{References}
\begin{thebibliography}{99}

\bibitem{zee}{\small A. Zee, {\itshape Quantum Field Theory in a
Nutshell} (Princeton University Press, 2003)}
\bibitem{klein1}{\small O. Klein, Z. Phys. {\bf 37}, 895
(1926)}
\bibitem{gordon}{\small W. Gordon, Z. Phys. {\bf 40}, 117
(1926)}
\bibitem{schrodinger1}{\small E. Schr\"odinger, Ann. Phys. (Berlin) {\bf
386}, 109 (1926)}
\bibitem{kaup}{\small D.J. Kaup, Phys. Rev. {\bf  172}, 1331 (1968)}
\bibitem{rb}{\small R. Ruffini, S. Bonazzola, Phys. Rev. {\bf  187}, 1767
(1969)}
\bibitem{colpi}{\small M. Colpi, S.L. Shapiro, I. Wasserman, Phys. Rev. Lett.
{\bf  57}, 2485 (1986)}
\bibitem{revueabril}{\small  A. Su\'arez, V.H. Robles, T. Matos, Astrophys.
Space Sci. Proc. {\bf 38}, 107 (2014)}
\bibitem{revueshapiro}{\small T. Rindler-Daller, P.R. Shapiro, Astrophys.
Space Sci. Proc. {\bf 38}, 163 (2014)}
\bibitem{bookspringer}{\small P.H. Chavanis, Self-gravitating Bose-Einstein
condensates, in {\it Quantum Aspects of Black Holes}, edited by X.
Calmet (Springer, 2015)}
\bibitem{mvm}{\small T. Matos, A. V\'azquez-Gonz\'alez,
J. Maga\~na, Mon. Not. R. Astron. Soc. {\bf 393}, 1359 (2009)}
\bibitem{abrilMNRAS}{\small A. Su\'arez, T. Matos, Mon. Not. R. Astron. Soc.
{\bf 416}, 87 (2011)}
\bibitem{abrilJCAP}{\small J. Maga\~na, T. Matos, A. Su\'arez,
F. J. S\'anchez-Salcedo, JCAP {\bf 10}, 003 (2012)}
\bibitem{khlopov}{\small M.Yu. Khlopov, B.A. Malomed, Ya.B. Zeldovich, Mon. Not.
R. astr. Soc. {\bf  215}, 575 (1985) }
\bibitem{chavaniscosmo}{\small P.H. Chavanis, Astron. Astrophys. {\bf 537}, A127
(2012)}
\bibitem{harkocosmo}{\small T. Harko, Mon. Not. R. Astron. Soc. {\bf 413}, 3095
(2011)}
\bibitem{shapiro}{\small B. Li, T. Rindler-Daller, P.R. Shapiro, Phys. Rev. D
{\bf 89}, 083536 (2014)}
\bibitem{madelung}{\small E. Madelung, Zeit. F. Phys. {\bf 40}, 322 (1927)}
\bibitem{bohmer}{\small C.G. B\"ohmer, T. Harko, J. Cosmol. Astropart. Phys.
{\bf 06}, 025 (2007)}
\bibitem{prd1}{\small P.H. Chavanis, Phys. Rev. D {\bf 84}, 043531 (2011)}
\bibitem{prd2}{\small P.H. Chavanis, L. Delfini, Phys. Rev. D {\bf 84}, 043532
(2011)}
\bibitem{chavharko}{\small P.H. Chavanis, T. Harko, Phys. Rev. D {\bf 86},
064011 (2012)}
\bibitem{smz}{\small  A. Su\'arez, T. Matos, Class. Quantum Grav. {\bf
31}, 045015 (2014)}
\bibitem{abrilph1}{\small  A. Su\'arez, P.H. Chavanis, Phys. Rev. D {\bf 92},
023510 (2015) }
\bibitem{abrilph2}{\small A. Su\'arez, P.H. Chavanis, in preparation}
\bibitem{ma}{\small C.-P. Ma, E. Bertschinger, Astrophys. J. {\bf
455}, 7 (1995)}
\bibitem{weinberg}{\small  S. Weinberg, {\it Gravitation and Cosmology} (John
Wiley, 1972)}
\bibitem{peebles}{\small P.J.E. Peebles, {\it The Large-Scale Structure of
the Universe} (Princeton University Press, 1980)}
\bibitem{bt}{\small J. Binney, S. Tremaine, {\it Galactic Dynamics} (Princeton
University Press, 2008)}
\bibitem{revuebec}{\small F. Dalfovo, S. Giorgini, L.P. Pitaevskii, S.
Stringari, Rev. Mod. Phys. {\bf 71}, 463 (1999)}
\bibitem{gh}{\small J.-A. Gu, W.-Y.P. Hwang, Phys. Lett. B {\bf 517}, 1 
(2001)}
\bibitem{widrow}{\small L.M. Widrow, N. Kaiser, Astrophys. J. Lett. {\bf 416},
L71 (1993)}
\bibitem{chavkpz}{\small P.H. Chavanis, Phys. Rev. D {\bf 84}, 063518 (2011)}
\bibitem{dvali}{\small G. Dvali, C. Gomez, Fortschr. Phys. {\bf 61}, 742 (2013)}
\bibitem{casadio}{\small R. Casadio, A. Orlandi, J. High Energy Phys. {\bf 8},
25 (2013)}
\bibitem{bl2}{\small C. Barcelo, S. Liberati, M. Visser, Phys. Rev. A
{\bfseries
68}, 053613 (2003)}
\bibitem{fock}{\small V. Fock, Z. Phys. {\bf 38}, 242
(1926)}
\bibitem{donder}{\small T. de Donder, H. van den Dungen, C. rend. Acad.
Sci. (Paris) {\bf 183}, 22 (1926)}
\bibitem{kudar}{\small J. Kudar, Ann. Phys. {\bf 81}, 632
(1926)}
\bibitem{dirac1}{\small P.A.M. Dirac, Proc. Royal Soc. A {\bf 117}, 610
(1928)}
\bibitem{dirac2}{\small P.A.M. Dirac, Proc. Royal Soc. A {\bf 126}, 360
(1930)}
\bibitem{anderson}{\small C. Anderson, Phys. Rev. {\bf 43}, 491
(1933)}
\bibitem{pw}{\small W. Pauli, V. Weisskopf, Helv. Phys. Acta  {\bf 7},
709 (1934)}



\end{thebibliography}
\end{document}